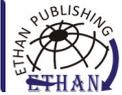

# Simulation of Switching Converters on the Level of Averaged Voltages and Currents

Aleksandra Lekić and Predrag Pejović

*School of Electrical Engineering, University of Belgrade, Belgrade 11120, Serbia*

**Corresponding author:** Aleksandra Lekić (lekic.aleksandra@etf.bg.ac.rs)

**Abstract:** An algorithm for simulation of switching converters is proposed in the paper. The algorithm is based on simulation of averaged circuit model applying "switching cell" concept, and construction of instantaneous values of the waveforms using quasi steady state and linear ripple approximation. Simulation covers converters operating both in the continuous and the discontinuous conduction mode. Application of the algorithm is demonstrated by simulation results of all three of the basic converters: buck, boost and buck-boost, as well as a flyback converter, which required slight generalization of the switching cell concept.

## 1. Introduction

Motivation for this research is in the fact that the most of available circuit simulators is general-purpose oriented [1] and apply unnecessarily complex device models, characterized by continuous and differentiable functions. On the other hand, the most important components of power electronics are switches, characterized by discontinuities in their constructive equations. Attempts to simulate power electronics circuits applying general-purpose simulation tools frequently yields convergence problems [2], reducing effectiveness of simulation in the design process, especially in cases when the converter control is focused. This motivated development of specialized simulation tools, intended to be applied in simulation of long term transients in power electronics circuits [2-4].

To avoid the convergence problems, an approach proposed here is based on simulation of averaged circuit model [5-6], with averaging applied on the "switching cell" level [7-9], and construction of actual waveforms, i.e., instantaneous values of the waveforms, applying quasi steady state approximation [10] and linear ripple approximation [5]. The algorithm presents an extension of the concepts proposed in [10] to cover discontinuous conduction mode and to generalize formulation of equations. Proposed approach results in a fast algorithm that provides instantaneous values of voltages and currents and their average values, minima, and maxima. Obtained waveforms could be post-processed further, aiming spectra of the waveforms, their RMS values, or any other waveform parameter.

## 2. Switching Cells

In this section, switching cell, used to describe both the CCM (Continuous Conduction Mode) and DCM (Discontinuous Conduction Mode) operating modes, is presented. Switching cells are treated as a three terminal devices. In this paper, two families of switching cells are described: For all three of the basic converters (buck, boost, and buck-boost) switching cells classified as Cell A family in [7] are used; for flyback converter, slightly modified switching cell that includes a transformer is used.

Two types of switching cells are considered for each family: Synchronous one that applies two bidirectional switches (state(S2) = ¬state(S1)), as well as the common type that applies a controlled switch and a diode. The converter with synchronous rectification always remains in CCM, while the converter with the diode might enter DCM. The switching cells are summarized in Table 1.

Averaged model of a switching cell used in simulations is shown in the Fig. 1. This model may represent a switching cell operating both in the continuous and in the discontinuous conduction mode, while preserving the circuit topological structure.

Focusing to a switching period $0 \leq t < T_S$, $T_S = 1/f_S$, and assuming that $i_{L0}$, $i_{L1}$ and $i_{L2}$ are inductor currents at the beginning of the period, $i_{L0} = i_L(0)$, at the moment $t = dT_S$, $i_{L1} = i_L(dT_S)$, and at the end of the switching period, $i_{L2} = i_L(T_S)$, using linear ripple approximation and quasi steady state approximation (as proposed in Ref. [10]), the set of equations describing basic switching cell given in Table 1 is formulated.

According to typical waveforms of $i_S$, $i_D$, and $v_L$ over a switching period, for both of the conduction modes expressions for $i_{L1}$, $i_{L2}$ and averages of currents of the switching devices and the inductor voltage are derived in the form

$$i_{L1} = i_{L0} + \frac{V_{L1}}{L} d T_S \qquad (1)$$

$$i_{L2} = i_{L1} + \frac{V_{L2}}{L} d_p T_S \qquad (2)$$

$$\overline{v_L} = d V_{L1} + d_p V_{L2} \qquad (3)$$

$$\overline{i_S} = \frac{d}{2}(i_{L0} + i_{L1}) \qquad (4)$$

$$\overline{i_D} = \frac{d_p}{2}(i_{L1} + i_{L2}) \qquad (5)$$

where overline indicates averaging over a switching period, $d_p = 1 - d$ in CCM, and $d_p = d_2$ in DCM, where $d_2$ is

$$d_2 = -\frac{V_{L1}}{V_{L2}} d \qquad (6)$$

The converter operates in the CCM if $d + d_2 \geq 1$ and in the DCM if $d + d_2 < 1$. The average model of the switching cell, consisting of Eqs. (1)-(6), covers both of the conduction modes, according to the accompanying inequalities. Eqs. (1)-(2) and (4)-(5) reduce to

$$\overline{i_S} = d i_{L0} + \frac{V_{L1}}{2L} d^2 T_S \qquad (7)$$

$$\overline{i_D} = d_p i_{L0} + \frac{V_{L1}}{L} d_p d T_S + \frac{V_{L2}}{2L} d_p^2 T_S \qquad (8)$$

which will be used to simulate the circuit. Current at the end of the interval is afterwards obtained as

$$i_{L2} = \frac{2\overline{i_D}}{d_p} - \frac{2\overline{i_S}}{d} + i_{L0} \qquad (9)$$

In a special case $d = 1$, $i_{L2} = i_{L1}$ and

$$i_{L2} = 2\overline{i_S} - i_{L0} \qquad (10)$$

while for $d_p = 1$

$$i_{L2} = 2\overline{i_D} - i_{L0} \qquad (11)$$

Switching cell for a flyback converter, presented in Table 1, is described by a slightly modified set of equations, due to the presence of a transformer. Furthermore, in comparison to original flyback converter, the cell is modified such that galvanic isolation is removed; in order to obtain connected graph. This is a requirement to perform numerical simulation, since numerical simulations can handle only circuits with unique solutions. It is still an open question whether this connecting should be performed at the switching cell level, or at the circuit level after the netlist is formed, which is primarily a matter of convenience of the user in specific applications.

**Table 1  Types of switching cells.**

|  | Containing bidirectional switches and the inductor | Containing switch, diode and inductor |
| --- | --- | --- |
| Basic cell | S1  S2  L | S  D  L |
| Flyback cell | S1  S2  1:n | S  1:n  D |

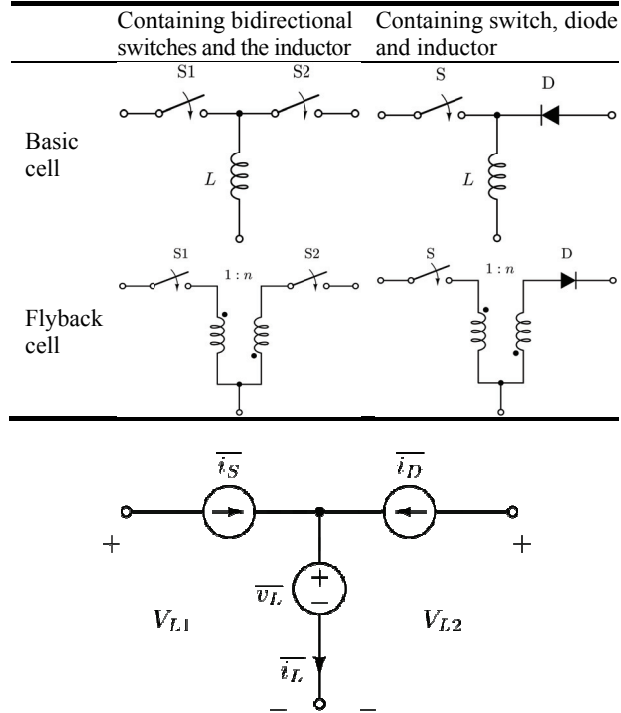

**Fig. 1  Averaged model of a switching cell.**

## 3. Forming Circuit Equations

To illustrate application of switching cells, an example of a buck converter, depicted in Fig. 2(a), is considered in this section. The example addresses synchronous buck converter, which always operates in the CCM. Identifying the switching cell of Fig. 1 in the buck converter circuit of Fig. 2(a) results in the circuit of Fig. 3, where the capacitor is replaced by its discretized equivalent according to the trapezoidal integration formula, consisting of branches 5 and 6 of the circuit. All of the currents and voltages in the circuit shown in Fig. 3 represent averaged values during $n^{th}$ switching period, for $nT_S \leq t < (n+1)T_S$.

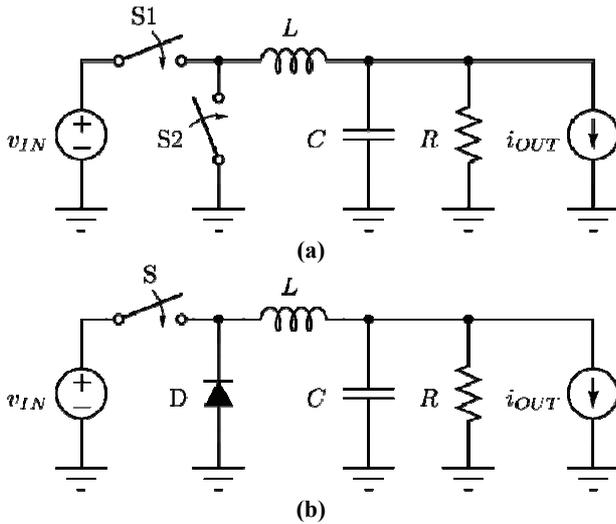

Fig. 2  The buck converter: (a) with synchronized bidirectional switches; (b) with a switch and a diode.

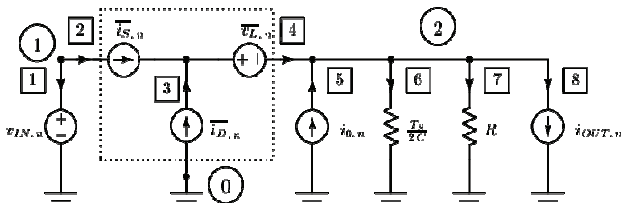

Fig. 3  Discretized averaged circuit model of the buck converter.

Discretized averaged circuit model of Fig. 3 is in each time step solved applying modified nodal analysis technique [1]. Other equations necessary to perform the simulation treat computation of independent sources in the circuit of Fig. 3, and besides the values of originally independent sources ($v_{IN}$ and $i_{OUT}$) arise from the trapezoidal integration formula and the equations that describe the switching cell. In the example of Fig. 3, the set of equations in a slightly optimized form of modified nodal analysis is given by

$$i_{1,n} + \overline{i_{S,n}} = 0 \qquad (12)$$

$$\left(\frac{2C}{T_S} + \frac{1}{R}\right)v_{2,n} - \overline{i_{S,n}} - \overline{i_{D,n}} = -i_{OUT,n} + i_{0,n} \qquad (13)$$

$$v_{1,n} = v_{IN,n} \qquad (14)$$

$$-\frac{v_{1,n}}{2L}d^2 T_S + \frac{v_{2,n}}{2L}d^2 T_S + \overline{i_{S,n}} = d\, i_{L0} \qquad (15)$$

$$-\frac{v_{1,n}}{L}d\, d_p T_S + \frac{v_{2,n}}{L}d_p T_S \left(d + \frac{d_p}{2}\right) + \overline{i_{D,n}} = d_p\, i_{L0} \qquad (16)$$

which is a linear system to be solved in each time step. Values of the switching cell current at the end of the period, $i_{L2}$, and the current source obtained as a result of the capacitor discretization, labeled as $i_0$, are updated. Value for $i_{L2}$ is obtained applying Eqs. (9)-(11). Independent source in the discretized representation of the capacitor is for the next switching interval given by

$$i_{0,n+1} = \frac{4C}{T_S}v_{2,n} - i_{0,n} \qquad (17)$$

Order of the optimized circuit equation system that describes the circuit is five, which corresponds to the circuit complexity. The system is used to demonstrate proposed approach.

## 4. Simulation Algorithm

Application of the algorithm is demonstrated by a simple program that simulates startup transient of the considered buck converter. The program input is a netlist containing elements of supported types, named SC for a switching cell, VDC for an independent voltage source, IDC for an independent current source, R for a resistor, and C for a capacitor. Names are followed by element indices. The element specification is followed by node numbers, three for a switching cell, and two for other elements supported this far, and the element parameter: inductance, voltage, current, resistance, or capacitance, depending on the element being specified. Specifications for a switching cell and a capacitor require another value to be specified to provide information about the initial value of the element current (for inductors in switching cells) or voltage (for capacitors). For the buck converter example of Fig. 2(a), the input file is

VDC 1 1 0 10.0
SCN1 1 1 0 2 10e-6 0
C 1 2 0 1e-4 0
R 1 2 0 5.0
IDC 1 2 0 4.0

*4.1. Formulation of Equations*

According to a given netlist that describes elements with their parameters and interconnections, the program forms arrays **A**, **x** and **z** that correspond to a linear equation system **Ax** = **z** of the discretized averaged circuit. Matrix **x** is a $(m + k + l) \times 1$ vector which holds the unknown variables. The first $m$ variables correspond to node voltages, $k$ elements in the middle represent currents of the voltage sources in the circuit, while the last $l$ variables are additional variables required by the switching cells. In the CCM of open loop systems, matrix **A** has constant coefficients, since the averaged circuit is linear, while in the DCM the averaged circuit is nonlinear, and some of the elements have to be adjusted in each iteration. Right hand side vector **z** consists of original independent sources, independent sources introduced by the discretization, and independent sources introduced by the switching cell model, the last two being dependent on the network solution in previous iteration.

In the case of the buck converter shown in Fig. 2, with the corresponding averaged and discretized model depicted in Fig. 3, described by Eqs. (12)-(16), the system that describes the circuit in $n^{th}$ switching period is given by

$$\begin{bmatrix} 0 & 0 & 1 & 1 & 0 \\ 0 & G_C + \frac{1}{R} & 0 & -1 & -1 \\ 1 & 0 & 0 & 0 & 0 \\ -\frac{d^2 G_L}{2} & \frac{d^2 G_L}{2} & 0 & 1 & 0 \\ -dd_p G_L & d_p G_L \left(d + \frac{d_p}{2}\right) & 0 & 0 & 1 \end{bmatrix} \begin{bmatrix} v_{1,n} \\ v_{2,n} \\ i_{1,n} \\ \overline{i_{S,n}} \\ \overline{i_{D,n}} \end{bmatrix} = \begin{bmatrix} 0 \\ -i_{OUT,n} + i_{0,n} \\ v_{IN,n} \\ di_{L0} \\ d_p i_{L0} \end{bmatrix} \quad (18)$$

where $G_C = (2C)/T_S$ and $G_L = T_S/L$. Solution of the circuit during $n^{th}$ switching period is

$$\mathbf{x}_n = \mathbf{A}_n^{-1} \mathbf{z}_n \quad (19)$$

Simulation of the averaged circuit involves the following steps: Forming the equation system (18), which corresponds to updating parameters of $\mathbf{A}_n$ and $\mathbf{z}_n$, computing $\mathbf{x}_n$, as well as the values of the inductor current at the end of the period and the current source obtained as a result by capacitor discretization for the next period.

The converter operating mode is checked in each time step and specified for the next step. The condition to resolve the operating mode is related to the inductor current, and assuming fixed voltages across the switching cell ports during a switching period and knowing the initial value of the inductor current, it can be predicted in advance to the simulation step. This approach limits the step size to correspond to the switching period.

After the values of vector **x** are computed in a number of time points in a considered interval, averaged waveforms of the inductor current and the capacitor voltage can be plotted, as a direct result of simulation. However, to construct instantaneous waveforms, quasi steady state approximation is used [10].

*4.2. Construction of Waveforms*

The first instantaneous value waveform to be constructed is the waveform of the inductor current. During each switching interval, the inductor current values at the beginning and at the end of the interval are known as direct results of the averaged circuit simulation. For the waveform construction in the continuous conduction mode, the only missing value is the current when the switching occurs, at $t = dT_S$, which can be obtained using known current at the beginning of the interval and the inductor voltage $V_{L1}$, as given by Eq. (1). In the case of the discontinuous conduction mode, an additional point where the inductor current reaches zero is required, $t = (d + d_p)T_S$, which is already computed.

Since the linear ripple approximation is applied, the instantaneous waveform of the inductor current is obtained by plotting these values and connecting them by straight line segments. In Fig. 4, inductor current waveforms are shown for the continuous and the discontinuous conduction mode.

In a similar manner, instantaneous voltage of the capacitor can be computed by superimposing the ripple to the average value, which will be discussed in the next section.

## 5. Simulation Results

To illustrate application of the algorithm, two examples are given in this section. The first example is a buck converter, which illustrates application of the basic switching cell, while the second example is a flyback converter, which illustrates application of the modified switching cell.

*5.1 Buck Converter*

In this example, two types of buck converters are used as benchmark circuits: a buck converter with synchronous rectification that applies switching cell depicted in Fig. 2(a), and a common buck converter that applies a controlled switch and a diode, depicted in Fig. 2(b). For both of the converters, startup transient is simulated. Expected average of the output voltage in the steady state is $V_{OUT} = 5$ V, and the inductor current average value is $I_L = 5$ A, which should be achieved at the end of simulation.

Simulation results are presented in Fig. 5, where the waveforms in the left column correspond to the converter with synchronous rectification, while the right column corresponds to the converter that applies a diode and might operate in the discontinuous conduction mode. In the diagrams of the inductor current, time intervals in which the converter operates in the DCM can be identified. Waveforms that correspond to the converter that enters the DCM in some time intervals expose faster convergence towards the steady state operation. Contribution of this paper in comparison to [10] is in enabling simulation of switching cells in discontinuous conduction modes.

In the example of synchronous buck converter shown in Fig. 2(a), the capacitor voltage ripple can be

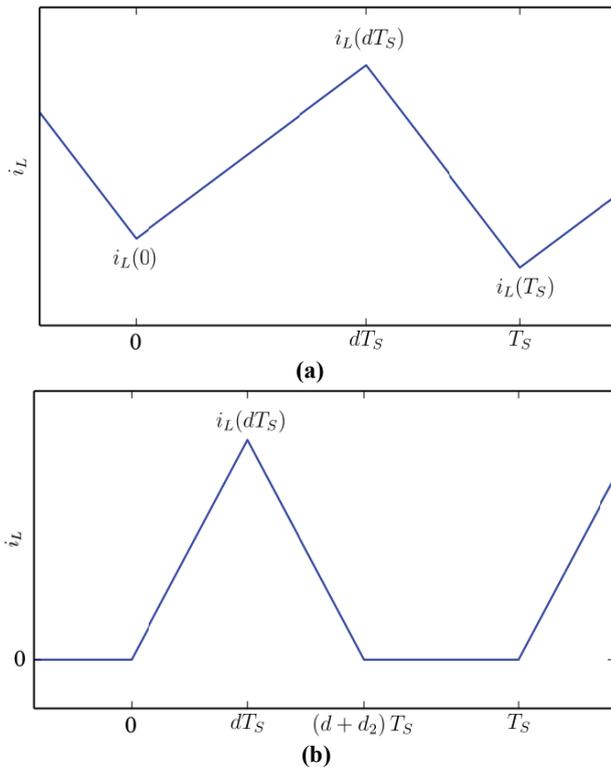

**Fig. 4  Inductor current waveforms: (a) continuous conduction mode; (b) discontinuous conduction mode.**

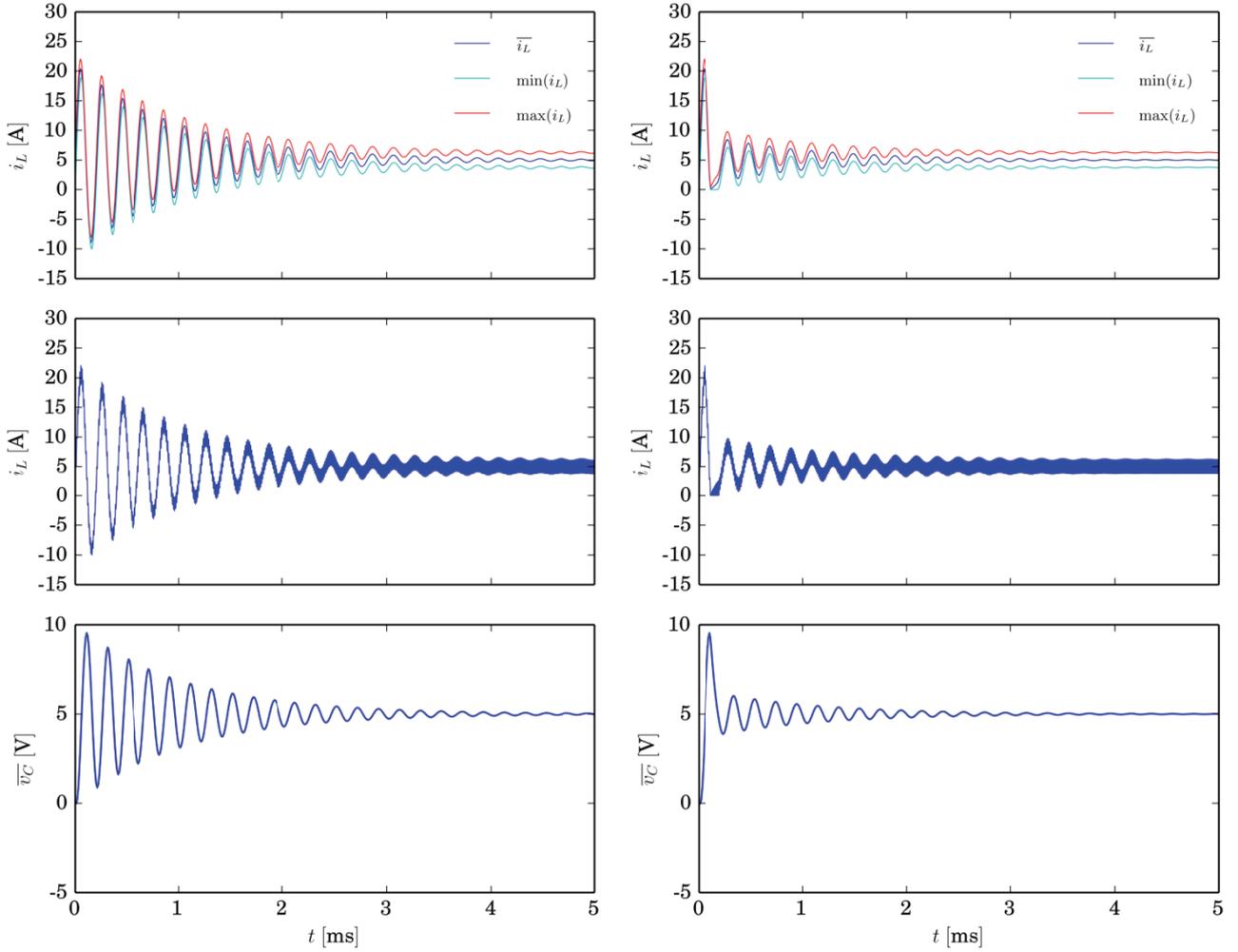

Fig. 5　Simulation results for the buck converter. Left column: converter of Fig. 2(a); right column: converter of Fig. 2(b). Parameters: $R = 5\ \Omega$, $I_{OUT} = 4$ A, $L = 10\ \mu$H, $C = 100\ \mu$F, $V_{IN} = 10$ V, $f_S = 100$ kHz, $D = 0.5$.

computed using the capacitor constitutive equation. In the converter, the capacitor current is equal to

$$i_C = C\frac{dv_C}{dt} = i_L - i_{OUT} - \frac{v_C}{R} \quad (20)$$

According to [10], the inductor current ripple can be determined as

$$\Delta i_{L1} = \frac{i_{L1} - i_{L0}}{2} \quad (21)$$

$$\Delta i_{L2} = \frac{i_{L1} - i_{L2}}{2} \quad (22)$$

$$\Delta i_L = \frac{\Delta i_{L1} + \Delta i_{L2}}{2} \quad (23)$$

Assuming that during the switching interval the inductor current ripple amplitude is as given by Eq. (23), the capacitor voltage ripple is determined by integrating Eq. (20). Applying the quasi steady state approximation again, the ampere-second balance is assumed to be approximately satisfied,

$$\overline{i_C} = \overline{i_L - i_{OUT} - \frac{v_C}{R}} = 0 \quad (24)$$

which means that the capacitor voltage variation during one switching period is primarily caused by the capacitor current ripple, not by its average value over the considered interval. The capacitor current ripple caused by the ripple of the resistor (R) current, in turn caused by the capacitor voltage ripple, is rather small in comparison to the inductor current ripple, thus neglected. The output current is assumed constant, and these facts remove the last two terms of Eq. (20) from the capacitor voltage ripple analysis. Therefore, the capacitor current ripple is equal to the inductor current ripple, $i_L - \overline{i_L}$. This results in an expression for the capacitor voltage

$$v_C(\tau) = \begin{cases} v_C(0) + \dfrac{\Delta i_L}{f_S C}\left(\dfrac{\tau^2}{d} - \tau\right) & 0 < \tau \leq d \\ v_C(0) + \dfrac{\Delta i_L}{f_S C}\dfrac{(\tau-d)(1-\tau)}{1-d} & d < \tau \leq 1 \end{cases} \quad (25)$$

where $\tau = t/T_S$, and $v_C(0)$ is the capacitor voltage at the beginning of the considered switching period. This voltage is equal to

$$v_C(0) = \overline{v_C} + \dfrac{2d-1}{6 f_S C}\Delta i_L \quad (26)$$

The capacitor voltage waveform for the buck converter operating in the continuous conduction mode is shown in Fig. 6.

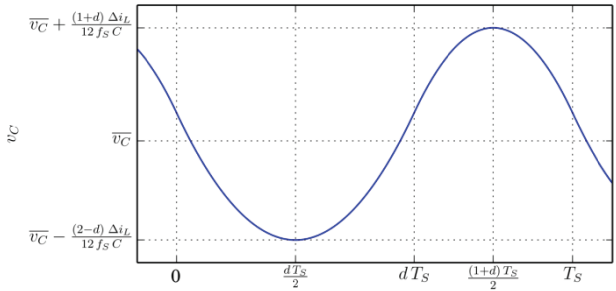

**Fig. 6 Capacitor voltage waveform in the continuous conduction mode.**

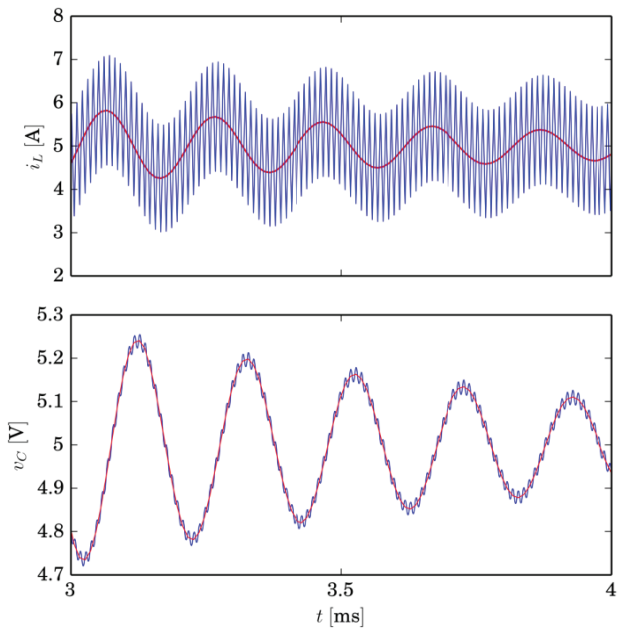

**Fig. 7 Simulation results of a buck converter of Fig. 2(a), inductor current and capacitor voltage. After superimposing ripple, blue trace; averaged value, red trace.**

Previously described simulation provides diagram of the average value of the capacitor voltage as given in Fig. 5. By linearly interpolating capacitor voltage value between the points obtained by the averaged circuit simulation, and superimposing the ripple computed according to Eqs. (25) and (26), the capacitor voltage waveform is constructed. In Fig. 7, inductor current and capacitor voltage waveforms for the synchronous buck converter of Fig. 2(a) are shown. In Fig. 7, transient waveforms in the time interval from 3 ms to 4 ms are plotted, where constructed instantaneous waveforms are depicted in blue and averaged values are red. This illustrates how the capacitor voltage ripple is included in simulation applying the same methods as in constructing the inductor current ripple.

### 5.2 Flyback Converter

To illustrate application of the flyback switching cell, a converter shown in Fig. 8(a) with synchronous rectification is used, as well as a common flyback converter depicted in Fig. 8(b). Simulation results are presented in Fig. 9. Expected average of the output voltage in the steady state, after the transient, is $V_{OUT} = 20$ V, while the inductor current average value in steady state is $I_L$ = 20 A. These values are obtained as the simulation result. Simulation of the transient covers the time interval of 5 ms. Since the switching frequency is assumed as 100 kHz, the waveforms are obtained for 500 switching periods. The waveforms of $i_L$ and $v_C$ are in agreement with the theoretical predictions.

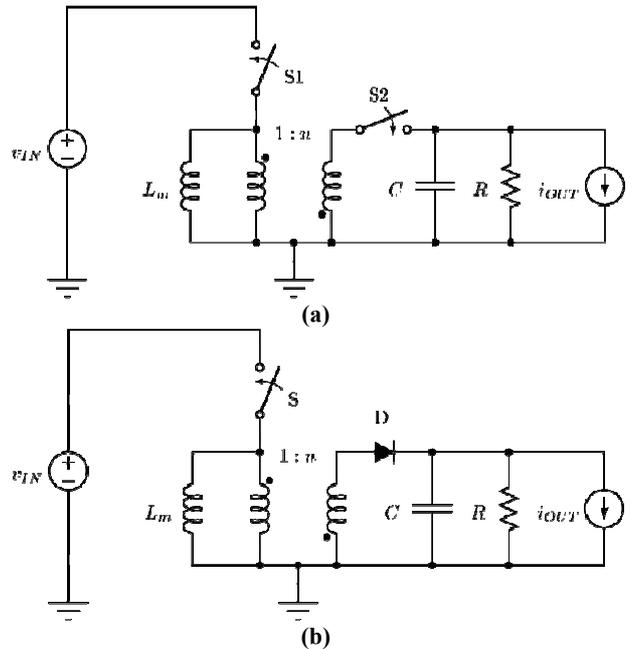

**Fig. 8 Simulated flyback converters: (a) with bidirectional switches; (b) with a switch and a diode.**

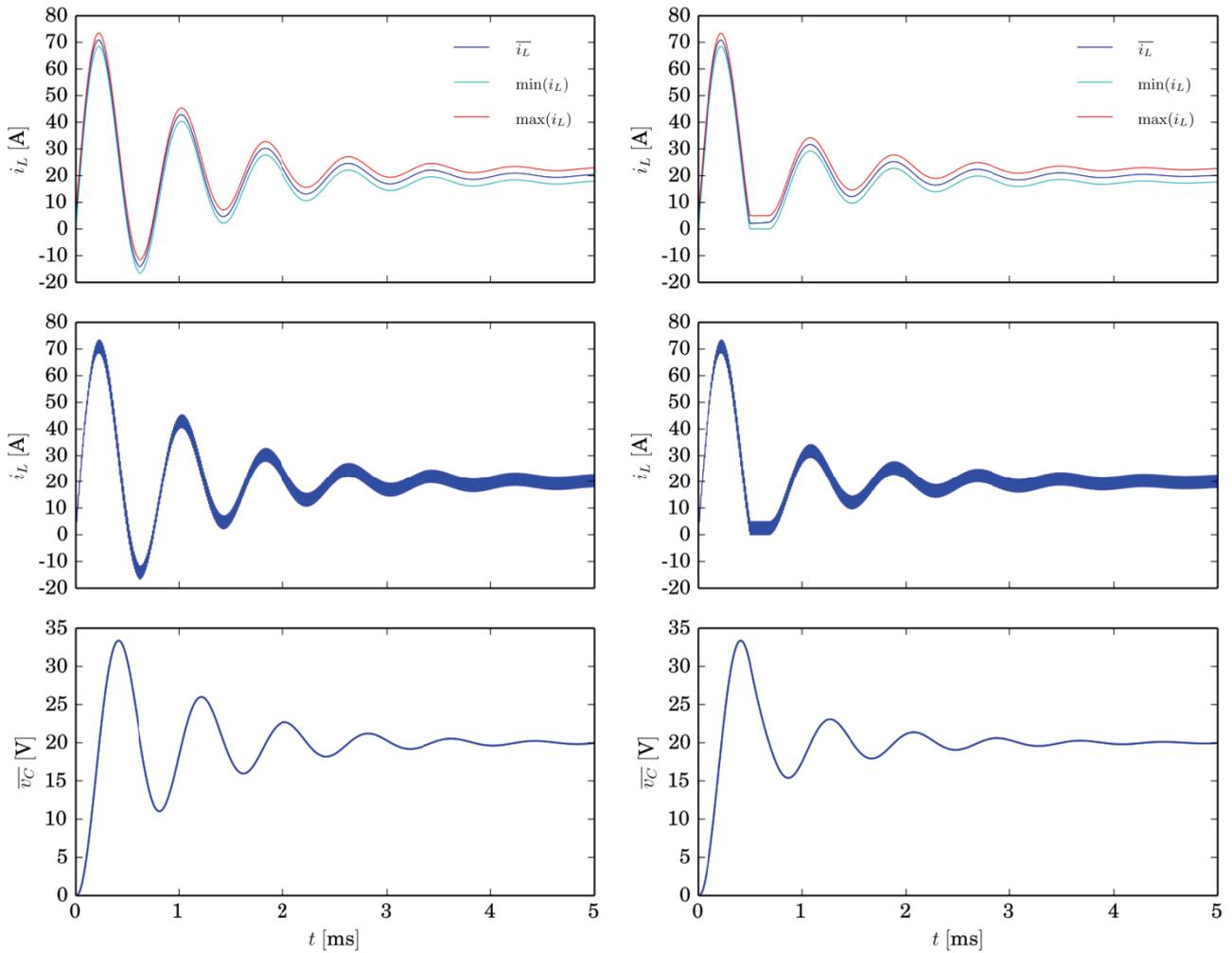

Fig. 9  Simulation results for the flyback converter. Left column: converter of Fig. 8(a); right column: converter of Fig. 8(b). Parameters: $R$ = 5 Ω, $I_{OUT}$ = 1 A, $L_m$ = 10 μH, $C$ = 100 μF, $V_{IN}$ = 10 V, $f_S$ = 100 kHz, $D$ = 0.5, $n$ = 2.

The simulation results are obtained by a program written specifically in this purpose, in Python 2.7.6 programming language, in PyLab environment, relying entirely on free software tools. The PyLab environment provides access to matplotlib package, which was used to plot the diagrams shown in Fig. 5 and Fig. 9. The numerical simulation alone, without generating the plots, was completed in 50 ms on a PC computer equipped with Intel Core 2 Duo P8400 processor run at 2.26 GHz under Ubuntu 14.04 operating system. It should be taken into account that the algorithm is implemented in Python programming language, which is interpretative, thus implementation in C should run much faster.

The program as an input requires a text file that contains a list of elements (capacitors, resistors, switching cells, etc.) and for a given duty ratio, switching frequency, transient duration, and the initial conditions, generates matrices **A**, **x** and **z**, and plots the inductor current and the capacitor voltage in the form given in Fig. 5 and Fig. 9.

## 6. Conclusions

An accurate and an efficient simulation algorithm for switching converters are proposed in this paper.

The algorithm performs simulation on the averaged circuit level applying switching cell concept. Convergence problems, as well as problems of determination of state of piecewise linear elements are avoided by the use of averaged circuit, the switching cell concept, the linear ripple approximation, and the quasi steady state approximation. As a result, a fast algorithm for averaged circuit simulation is obtained with the possibility to reconstruct waveforms of actual circuit currents and voltages applying linear ripple and quasi steady state approximations. Both the continuous and the discontinuous conduction mode are handled by the proposed method, primarily due to the application of the switching cell concept. Application of the switching cell concept is an improvement in comparison to [10], as well as the support for the discontinuous conduction mode.

Proposed algorithm is implemented in a program written in Python programming language. Regardless the fact that Python is an interpretive language, simulation is really fast, taking only 50 ms to simulate 5 ms transient. This indicates that proposed method is promising to become a useful and an efficient tool for power electronic circuits. Further development is directed toward formalizing forming of the equation system, formalizing construction of the instantaneous waveforms on the basis of obtained averaged waveforms, analysis of accuracy of applied numerical integration method, and application of the method for true nonlinear circuits, like the converters with the feedback control system applied.